\documentclass[twocolumn]{aastex63}
\usepackage{graphicx}
\usepackage{bm}
\usepackage{color}
\usepackage{soul}
\usepackage{hyperref}
\usepackage{enumerate}
\usepackage{ulem}

\newcommand{\beq}{\begin{equation}}
\newcommand{\eeq}{\end{equation}}
\newcommand{\beqn}{\begin{eqnarray}}
\newcommand{\eeqn}{\end{eqnarray}}

\newcommand{\maxtov}{M_{\rm max}^{\rm sph}}
\newcommand{\maxsup}{M_{\rm max}^{\rm sup}}

\begin{document}

\title{GW190814: Spin and Equation of State of a Neutron Star Companion}

\author{Antonios Tsokaros}
\affiliation{Department of Physics, University of Illinois at
  Urbana-Champaign, Urbana, IL 61801, USA}    

\author{Milton Ruiz}
\affiliation{Department of Physics, University of Illinois at
  Urbana-Champaign, Urbana, IL 61801, USA}

\author{Stuart L. Shapiro}
\affiliation{Department of Physics, University of Illinois at
  Urbana-Champaign, Urbana, IL 61801, USA}
\affiliation{Department of Astronomy \& NCSA, University of
  Illinois at Urbana-Champaign, Urbana, IL 61801, USA}

\correspondingauthor{Antonios Tsokaros}
\email{tsokaros@illinois.edu}

\date{\today}

\begin{abstract}
The recent discovery by LIGO/Virgo of a merging binary having a $\sim 23 M_\odot$ black hole and a
$\sim 2.6 M_\odot$ compact companion has triggered a debate regarding the nature of the secondary, which
falls into the so-called mass gap. Here we explore some consequences of the assumption that the
secondary was a neutron star (NS). We show with concrete examples of heretofore viable equations of
state (EOSs) 
that rapid uniform rotation may neither be necessary for some EOSs nor sufficient for
others to explain the presence of an NS. Absolute upper limits for the maximum mass of a spherical
NS derived from GW170817 already suggest that this unknown compact companion might be a
slowly or even a nonrotating NS. However, several soft NS EOSs favored by GW170817 with maximum
spherical masses $\lesssim 2.1 M_\odot$ cannot be invoked to explain this object, even allowing for
maximum uniform rotation. By contrast, sufficiently stiff EOSs that yield $2.6 M_\odot$ NSs that are
slowly rotating or, in some cases, nonrotating, and are compatible with GW170817 and the
results of the Neutron Star Interior Composition Explorer (NICER), can account for the black hole companion.
\end{abstract}




\section{Introduction}

On 2019 August 14 the LIGO/Virgo Scientific Collaboration (LVC) made one of the most
intriguing gravitational-wave detections to date \citep{Abbott:2020khf}. The 
designated event,
GW190814, involved a binary coalescence that had the most asymmetric mass ratio
to date, $0.112_{-0.009}^{+0.008}$. The binary contained a primary component 
with mass $m_1=23.2_{-1.0}^{+1.1} M_\odot$ and dimensionless spin $\chi_1\leq 0.07$, 
presumably making it a very low-spinning black hole (BH). 
The mass of the secondary was $m_2=2.59_{-0.09}^{+0.08}$, placing it at 
the boundary of the so-called
``mass gap" \citep{Bailyn_1998,Ozel_2010} and therefore making its 
identification difficult 
\citep{Hannam_2013,Littenberg:2015tpa,Mandel:2015spa,Yang_2018}. The absence of an electromagnetic
counterpart and measurable tidal deformation add further uncertainty to 
the nature of this compact object and allows the secondary to be 
a BH, an NS, or something more exotic. 
Preliminary arguments based on estimates of the maximum spherical mass 
of an NS, the Tolman-Oppenheimer-Volkoff (TOV) limit, suggest that the 
unknown compact object was too heavy to be an NS \citep{Abbott:2020khf}. 

The TOV limit, $\maxtov$, is associated with the ground state of matter at 
zero temperature.
Setting constraints on $\maxtov$ is a long-standing pursuit 
(see reviews by
\cite{2016PhR...621..127L,Oertel2017,Baym2018})
but the detection of a low-mass binary coalescence, GW170817, by 
LIGO/Virgo  \citep{TheLIGOScientific:2017qsa} 
has played a significant role recently.
This binary system had total mass of $2.74^{+0.04}_{-0.01}\, M_\odot$ 
or $2.82_{-0.09}^{+0.47}\, M_\odot$ depending on the assumed priors for 
its dimensionless spin. The low mass estimate
assumed that the NSs had spin $|\chi|\leq 0.05$, while the 
high mass estimate assumed spin $|\chi|\leq 0.89$. 
Coincident with the detection of the gravitational waves and $1.734\pm\,0.054\,$s 
after the GW170818 inferred binary coalescence time, there was a 
short $\gamma$-ray burst, GRB 170817A, of duration 
$2\pm 0.5\,\rm s$ detected by the Fermi Gamma-Ray Burst Monitor~\citep{2017GCN.21520....1V, 2017GCN.21517....1K} and 
INTEGRAL~\citep{Savchenko:2017ffs,Savchenko17GCN}.

Following event GW170817 a large number of studies appeared in an effort to elucidate the properties of  
NSs and their supranuclear density regime
\citep{Shibata2017,Margalit2017,Bauswein_2017,Ruiz:2017due, Rezzolla_2018,Annala2018,Radice_2018,Most2018,De2018,Abbott2018,
Raithel_2018,Tews2018,Malik2018,Landry2019,Baym:2019iky,Shibata2019,Abbott_2020}.
Quantities rigorously investigated included the NS radius and its tidal deformability. 
For example, \cite{Tews2018} employed models constrained by calculations of 
the neutron-matter EOS, which employed
chiral effective-field theory Hamiltonians to predict that 
a $1.4M_\odot$ NS must have a radius $9.0\ {\rm km}< R_{1.4} < 13.6\ {\rm km}$, similar to \cite{Annala2018}. 
In addition they showed that NSs with $\maxtov \sim 2.5 M_\odot$ could be possible candidates for GW170817.

Independent of the LIGO/Virgo results, NS properties have been reported
recently by the Neutron Star Interior Composition Explorer (NICER) team. 
In particular, estimates of the mass and radius
of the isolated $205.53$ Hz millisecond pulsar PSR J0030+0451  were obtained using a
Bayesian inference approach to analyze its energy-dependent thermal X-ray waveform. 
It was shown that $R=13.02^{+1.24}_{-1.06}$ km and $M=1.44^{+0.15}_{-0.14}\ M_\odot$
\citep{Miller_2019,Riley_2019}, which indicate a stiffer EOS than
those mostly favored by the LVC.

Since the detection of GW190814 \citep{Abbott:2020khf} one scenario that could explain the BH 
companion was that of a rapidly spinning NS. Well-known studies by
\cite{Cook:1993qr,Cook:1993qj} were the first to show that spinning up an NS uniformly 
can increase its mass by $\sim 20\%$.
Therefore, uniform rotation could provide a means of explaining a heavier compact object, at least in
principle. This scenario also has been proposed by \cite{Most:2020bba}, who estimated the dimensionless
spin of the secondary to be $0.49\lesssim  \chi \lesssim  0.68$. Here we further consider the idea of an
NS as the black hole companion of GW190814 and explore its consequences. 
By employing concrete examples we show that rapid uniform rotation
may neither be necessary nor sufficient to explain the presence of
a $2.6 M_\odot$ NS in GW190814.
Soft EOSs consistent with GW170817, such as SLy \citep{Douchin01}
are unable to provide enough mass to explain the secondary in GW190814, 
even for an NS endowed with maximum uniform rotation. 
On the other hand, we argue that well-known 
absolute mass upper limits derived from GW170817 can be invoked to show that
a slowly or even a nonrotating NS can account for the secondary
for viable stiff EOSs, such as DD2 \citep{HEMPEL2010210}.
To put it differently, there are \textit{two ways} to connect events GW170817
and GW190814, and the scenario that the secondary object of the latter is an NS:
The first way is to assume low spin priors for GW170817, which then would imply
that the NS in GW190814 must be rapidly rotating and supported by a soft EOS
like SLy. In the second way one can assume high spin priors in GW170817, which
then can explain the secondary in GW190814 as a slowly or even nonrotating NS
supported by a stiff EOS like DD2.


\section{Assumptions}
\label{sec:ass}

In this work we assume that the companion of the BH in event GW190814
is a \textit{uniformly} rotating NS, i.e., an NS rotating 
at a frequency below the mass-shedding (Keplerian) limit, whose maximum mass 
we will denote by $\maxsup$. Rotating NSs beyond that
limit are called {\it hypermassive} (HMNS; \cite{BaShSh}) and are supported in
part by differential rotation. HMNSs are transient objects 
that typically collapse to a BH on timescales of $10-1000$ ms 
due to the redistribution or loss of angular momentum by 
viscosity, magnetic field winding and turbulence, and/or (if nonaxisymmetric) gravitational waves.
Uniformly rotating NSs with mass less than the Keplerian 
limit $\maxsup$ but larger than the maximum spherical limit, $\maxtov$,
are called {\it supramassive} \citep{CookShapTeuk} and may also eventually 
collapse to BHs
due to magnetic dipole radiation or gravitational waves, 
but on much longer timescales. If the 
rotating star has mass less than the maximum spherical limit it will 
remain forever as an NS.

In principle our analysis can hold true even for hybrid 
(nuclear plus quark matter) stars \citep{Paschalidis:2017qmb}.

\begin{figure*}
\includegraphics[width=0.49\textwidth]{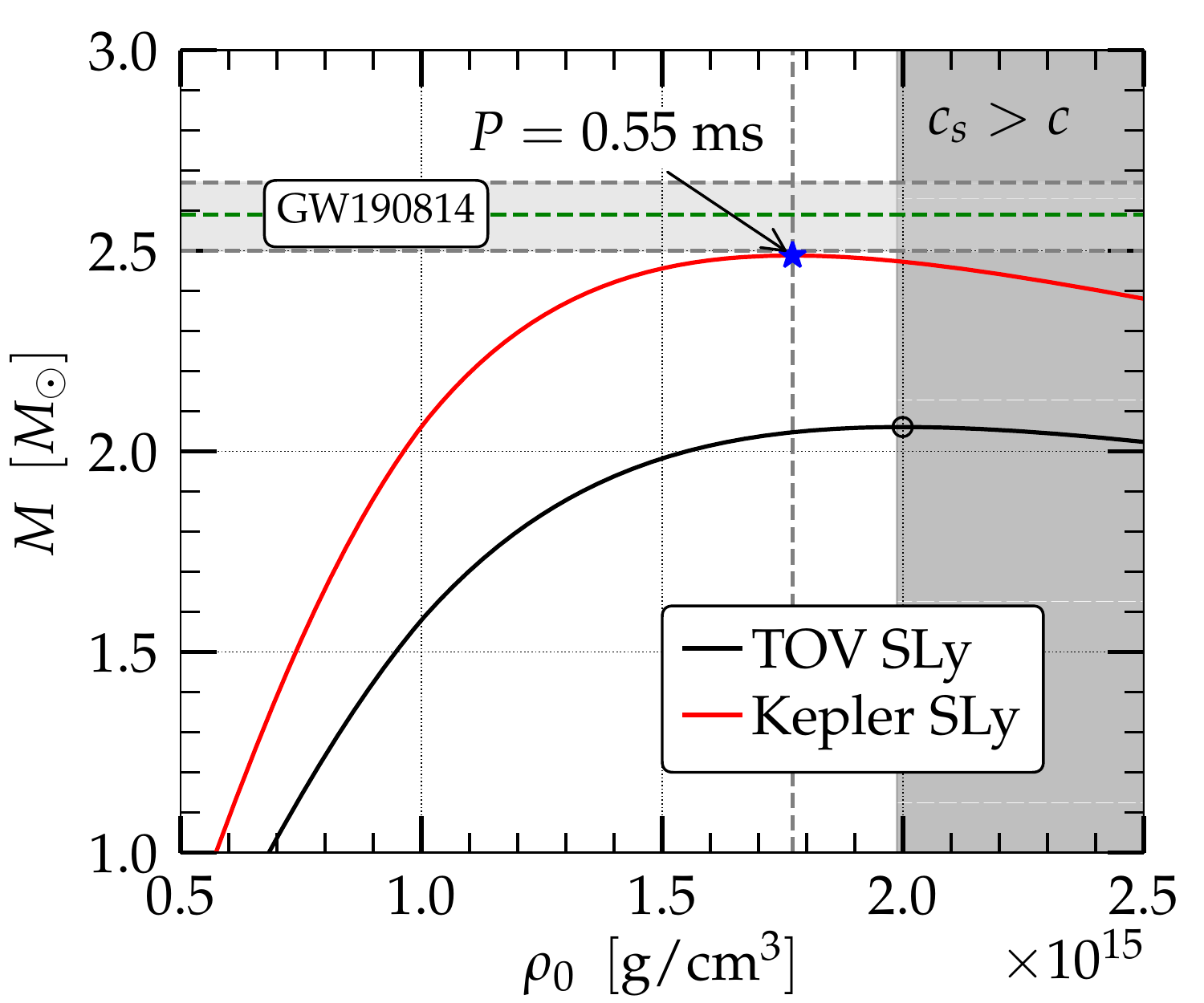}
\includegraphics[width=0.49\textwidth]{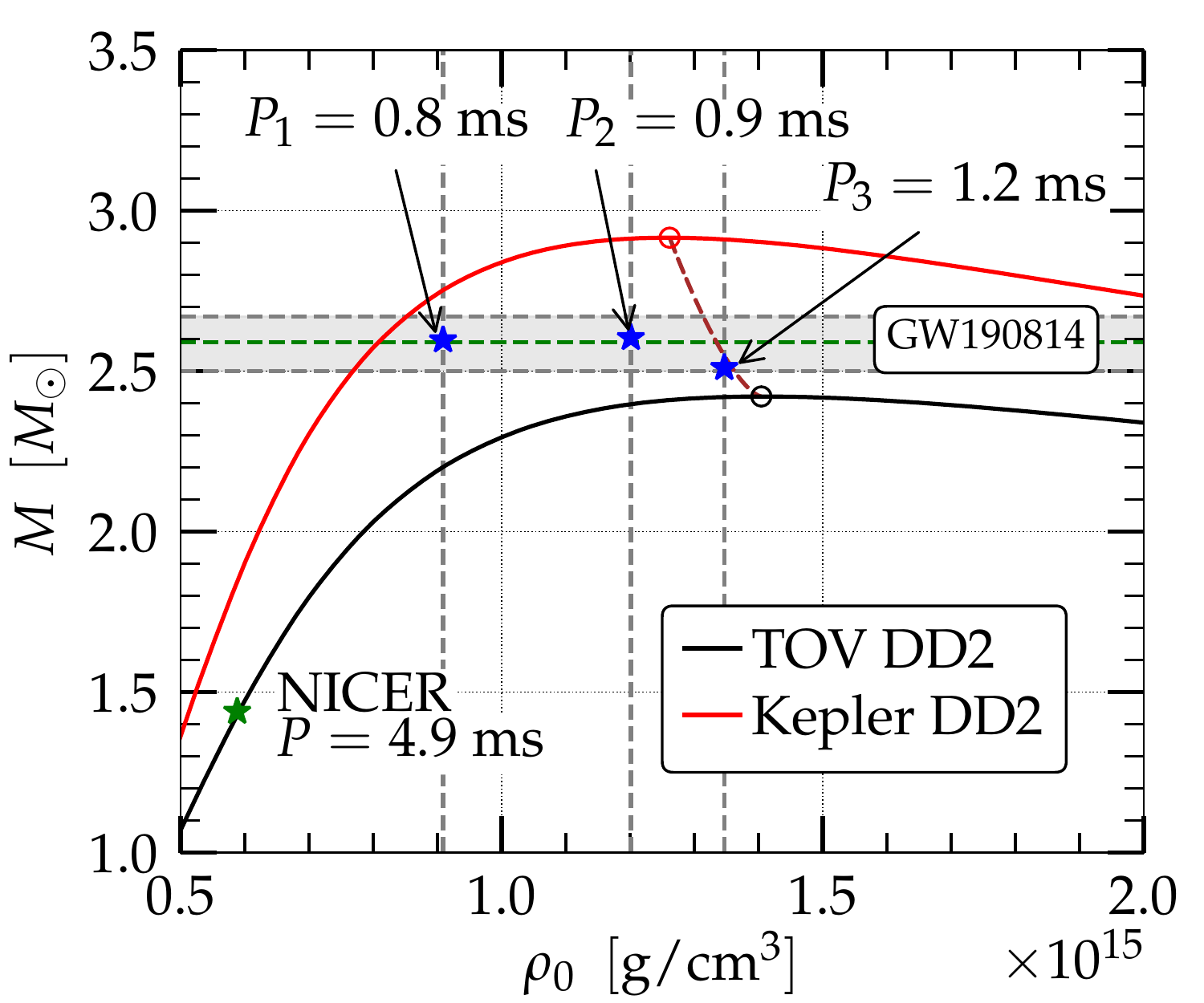}
\caption{Two possibilities for the EOS of an NS 
secondary in GW190814. 
The left panel employs the soft SLy EOS and fails to provide a 
model of a viable uniformly rotating star.
By contrast, the right panel employs the moderately stiff DD2 EOS 
and is successful.}
\label{fig:per}
\end{figure*}


\section{GW170817 and the maximum mass}
\label{sec:maxmass}

The GW170817 detection has triggered different techniques to estimate
$\maxtov$.
Based on information inferred from the 
electromagnetic and gravitational-wave spectra 
\cite{Margalit2017} conclude that 
$\maxtov \leq 2.17 M_\odot$ at $90\%$ confidence. They also argue that most probably
the remnant was an HMNS or a very short-lived supramassive remnant.
Using different arguments for the kilonova, together with quasi-universal 
relations between $\maxtov$ and the supramassive mass limit
$\maxsup$, \cite{Rezzolla_2018} found  
$2.01^{+0.04}_{-0.04}\ M_\odot \leq \maxtov \leq 2.16^{+0.17}_{-0.15}\ M_\odot$, yielding an absolute 
upper bound of $\maxtov \leq 2.33\ M_\odot$. 
They assumed that the core collapsed exactly
at the maximum mass-shedding limit. Another set of studies by \cite{Shibata2017,Shibata2019} concluded 
that this upper limit can be only weakly constrained to
be $\maxtov \lesssim 2.3 M_\odot$. 

Based on numerical GRMHD simulations \cite{Ruiz:2017due} have shown that the event GW170817 and its associated
short $\gamma$-ray burst GRB 170817A can be explained if the remnant object 
was a, HMNS, i.e. whereby
\begin{equation}
\beta \maxtov \approx \maxsup \lesssim M_{\rm GW170817} \lesssim  M_{\rm thresh} \approx \alpha \maxtov
\label{eq:gw170817}
\end{equation}
where $\alpha \approx 1.3 - 1.7$ is the ratio of the 
HMNS threshold mass limit (the limit that distinguishes prompt versus delayed collapse for the postmerger object) 
to the NS spherical maximum mass as calculated by various
numerical experiments \citep{Shi05,Shibata:2006nm,Baiotti:2008ra,Hotokezaka:2011dh,Bauswein:2013jpa}.
The important factor that bounds the NS mass from above is the $\beta$ parameter, which for
different realistic EOSs has been found to be $\beta \approx 1.20$ 
\citep{Cook:1993qj,Cook:1993qr,Lasota:1995eu,Breu:2016ufb},
while a general Rhoades-Ruffini causality argument 
yields $\beta \approx 1.27$ \citep{1987ApJ...314..594F,1997ApJ...488..799K}. 
Depending on the mass of the remnant and the $\beta$ parameter we were able to set upper limits on $\maxtov$
\citep{Ruiz:2017due} as follows:

Low-spin priors, $|\chi|\leq 0.05$,
\begin{equation}
\maxtov \lesssim \left\{ 
\begin{array}{ll}
2.16\pm 0.23 & \textrm{if $\beta\approx 1.27$} \\
2.28\pm 0.23 & \textrm{if $\beta\approx 1.20$}
\end{array} \right.
\label{eq:maxlow}
\end{equation}
while for high-spin priors, $|\chi|\leq 0.89$, 
\begin{equation}
\maxtov \lesssim \left\{ 
\begin{array}{ll}
2.22\pm 0.66 & \textrm{if $\beta\approx 1.27$} \\
2.35\pm 0.66 & \textrm{if $\beta\approx 1.20$}
\end{array} \right.  .
\label{eq:maxhigh}
\end{equation}

A hard lower bound on $\maxtov$ is set by measurements of pulsar masses, which
to date are $2.01^{+0.04}_{-0.04} M_\odot$ for J0348+0432 \citep{2013Sci...340..448A}, 
$1.928^{+0.017}_{-0.017} M_\odot$ for J1614-2230 \citep{2010Natur.467.1081D}, and
J0740+6620 $2.14^{+0.20}_{-0.18} M_\odot$  \citep{Cromartie:2019kug}
These measurements suggest that $\maxtov\gtrsim 2.0 M_\odot$ while
equations (\ref{eq:maxlow}) and (\ref{eq:maxhigh}) suggest that the absolute limit for the
maximum mass of spherical NSs, for the majority of the EOSs ($\beta\approx 1.20$),
is $\maxtov\lesssim 2.51$ if GW170817 was composed of low-spin NSs and $\maxtov\lesssim 3.01$ if it was
composed by high-spin ones. These absolute NS upper limits can be invoked 
already to suggest a straightforward explanation for ``heavy'' compact 
objects like the ones in GW190814 or GW190425 
\citep{Abbott_2020,Abbott:2020khf} without having to identify specific
nuclear models or, more significantly, resort to extreme physics.


\section{Consequences for the EOS of GW190814}
\label{sec:gw190814}

To assess the possibility that GW190814 contains a uniformly rotating NS a choice for an EOS 
necessarily has to be made. Although the correct EOS that describes supranuclear densities is not 
currently known, here we choose the SLy \citep{Douchin01} and the DD2 \citep{HEMPEL2010210} EOSs.
These EOSs, which are broadly compatible with GW170817 data, are chosen not because they are 
more viable than others but rather because they exhibit ``opposite'' behaviors that will 
help us illustrate a point.
In particular, the low-spin prior prediction for the GW170817 mass is 
$2.73^{+0.05}_{-0.01}M_\odot$
\citep{GW170817_new},\footnote{
Here we report the TaylorF2 values. The estimates from SEOBNRT and PhenomDNRT \citep{GW170817_new}
are similar, and do not change the argument we put forward.  }
a value that is easily accommodated by a transient HMNS with the SLy EOS. 
On the other hand, the high-spin prior mass prediction of GW170817, $2.79^{+0.30}_{-0.06}M_\odot$,
is closer to the prompt collapse threshold for the SLy EOS, $\sim 2.9 M_\odot$ \citep{Bauswein:2013jpa}.
Binary NS simulations suggest that the lifetime of the HMNS remnant close to the prompt collapse limit
is smaller than its lifetime when it is close to the supramassive limit.
Given the fact that the HMNS remnant of GW170817 must have survived for $\sim 1\rm s$ \citep{Gill:2019bvq}
before collapse to a BH, the high-spin prior case ($2.79M_\odot$ remnant mass) is not as probable as the 
low-spin one ($2.73M_\odot$ remnant mass) for the SLy EOS.
This picture differs significantly in the DD2 EOS. In fact, DD2 is incompatible with the low-spin 
prior estimate of GW170817 and the requirement of Equation (\ref{eq:gw170817}) that it be hypermassive, 
given that $\maxsup=2.92 M_\odot$. Assuming instead high spins for GW170817, Equation (\ref{eq:gw170817}) now requires 
$\maxsup\lesssim 3.09 M_\odot$ ($(2.79+0.30)M_\odot$),  which is compatible with the $\maxsup$ of DD2.

In Figure \ref{fig:per} we plot the mass versus rest-mass density for the 
SLy \citep{Douchin01} and the DD2 \citep{HEMPEL2010210} EOSs. 
In these plots we denote by a black solid line spherical, TOV NS 
models, while with a red solid line we show models at the mass-shedding 
(Kepler) limit.
The models were computed using the relativistic rotating equilibrium code of \cite{Cook:1993qr,Cook:1993qj}.
The maximum of these
curves represents $\maxtov$ and $\maxsup$, respectively. 
No \textit{uniformly} rotating star can exist
above the red lines in Figure \ref{fig:per}. This means that the compact object in GW190814
\textit{cannot be explained by an EOS like SLy}, although it is 
favored by the event GW170817 \citep{Abbott:2018exr}. Such EOSs must now be rejected 
because their mass-shedding limit is below the lower limit mass of the
compact object in GW190814, i.e.,
\begin{equation}
 \maxsup < 2.59^{+0.08}_{-0.09} \ M_\odot
\label{eq:sly}
\end{equation}
These EOSs they all share a relatively low maximum spherical mass, $\maxtov \lesssim 2.1$, and thus they 
are called {\it soft} (see \cite{Read:2008iy} for a list of various mass limits). 
Notice the supramassive limit 
for SLy, denoted by
a blue star in Figure \ref{fig:per}, has a dimensionless spin of $\chi=0.7$ and a rotational period $P=0.55$ ms. 
Hence \textit{the fact that the NS is rapidly spinning and reaches a high $\chi$ does not mean that it 
can necessarily explain GW190814}.  
A glance at \cite{Read:2008iy} reveals that a great number of these soft 
EOSs are ruled out based on this simple observation. 
In addition, the model at the supramassive limit, is both dynamically and secularly unstable 
since both the dynamical and secular stability points reside slightly to the left of the turning point 
at lower rest-mass densities \citep{2011MNRAS.416L...1T,1988ApJ...325..722F}.

In light of GW170817 the above conclusions can be interpreted in either of 
two ways.
One way suggests that there is a tension between the EOSs that 
are favored by GW170817 and those that can 
be used to explain the NS companion in GW190814. 
This does not imply that \textit{all} EOSs favored 
by GW170817 are nonviable candidates for GW190814, but there is certainly a 
gap between the two sets.
The other way acknowledges that GW170817 is a binary NS system 
subject to well-established EOS restrictions that lead one to favor
certain EOSs, 
while the nature of the secondary in GW190814 is uncertain.
Hence, we might conclude that the likelihood of the secondary in GW190814 
being a rotating NS may be small.

A second scenario is depicted in the right plot of Figure \ref{fig:per} where 
we invoke the DD2 EOS to represent a different class of models. 
Here the GW190814 limits are easily accommodated within the supramassive regime 
$[\maxtov,\maxsup]$. In the spirit of Equation (\ref{eq:gw170817}),  
for the GW190814 secondary to be a uniformly rotating NS we must have
\begin{equation}
  2.59^{+0.08}_{-0.09} \ M_\odot \ \lesssim \ \maxsup\  \approx \beta \maxtov 
\label{eq:aa}
\end{equation}
which for $\beta\sim 1.20$ gives immediately $\maxtov \gtrsim 2.1 M_\odot$, 
consistent with modern pulsar 
observations \citep{Cromartie:2019kug}.
In \cite{Most:2020bba} the same bound was found
using more complicated arguments based on universal relations 
emerging from numerical fits.
Three models are depicted with blue stars having periods around $1$ ms. All of them reside on the left of
the turning point line (maxima on constant angular momentum curves) 
depicted by a brown dashed line and therefore are dynamically and secularly stable with respect to 
axisymmetric perturbations, which include radial modes 
\citep{1988ApJ...325..722F,2011MNRAS.416L...1T}. 
In addition they are all dynamically and secularly
stable with respect to nonaxisymmetric $m=2$ (bar) modes (all models have $T/W\leq 0.1$ where $T$ is the 
rotational kinetic and $W$ is the gravitational binding energy). Although these
stars are highly rotating from an astrophysical point of view, not all of them are considered rapidly rotating
NSs in a general relativistic context. In particular 
the third model with period $P_3=1.2$ ms
has $\chi_3=0.34$,  $T/W=0.03$ and deformation
$R_p/R_e=0.91$, where $R_p$ and $R_e$ are the polar and equatorial radii, 
respectively. In general relativity
this model is considered a slowly rotating star for which even the slow-rotation 
approximation for equilibria \citep{Hartle1967} can provide an accurate description. 
By contrast, the model with maximum
supramassive mass at the Keplerian limit (shown with a red circle) has 
$\chi=0.7$,  $T/W=0.13$,  $R_p/R_e=0.56$, and $P=0.7$ ms. 

We note that while the periods quoted above may be short astrophysically,
they are not unduly so. The fastest-spinning observed pulsar PSR J1748-2446ad
has a period of $\sim 1.4$ ms \citep{2006Sci...311.1901H}, which resides 
in the neighborhood of the above values. 
Short periods are consistent with the requirement that pulsars must have 
sufficiently small exterior B-fields to avoid spin-down over a 
reasonable lifetime. This is typically the case for recycled pulsars. 
Small fields generate small electromagnetic dipole emission and, if the 
radio luminosity is correspondingly low, this may explain why 
we have not observed the most rapidly rotating NSs with periods below 
$\sim$ millisecond as radio pulsars.
Regarding compact binary systems, approximately 20 binary NSs have been detected
\citep{Tauris:2017omb,Zhu:2017znf} while there are no robust detections of a
binary BH-NS so far. The NS in J1807-2500B has a period of $4.2\rm ms$, or 
$\chi \sim 0.12$,
while  others typically have longer periods  (smaller $\chi$). While the above set of
observations is small and one cannot draw definitive conclusions, one might safely
argue that if spin-down due to electromagnetic emission is as efficient as in the currently known
binaries, then any scenario involving a highly spinning NS either in a binary NS or
in a binary BH-NS system is not probable. Finally, for these reasons and others 
we also note that it has been argued that 
GW190814 is more likely a binary BH \citep{Abbott:2020khf}.

The major point of the right panel of Figure \ref{fig:per} is 
demonstrating with a concrete example 
that {\it with a relatively stiff EOS}
\citep{Tews2018,Annala2018,Tan:2020ics,Alsing:2017bbc}
{\it the secondary in GW190814 can be a slowly rotating NS.} 
Moreover, {\it if the EOS was only slightly
stiffer, the secondary could even be a nonrotating companion.} 
Indeed, GW170817 and the maximum mass limits that it has spawned 
(section \ref{sec:gw190814}) show that
a slowly or even nonrotating NS for the secondary in GW190814 can 
be realized in principle. 
For nonrotating priors in GW170817 the absolute upper limit for $\maxtov$ is $2.51$, while for highly spinning 
priors it is $3.01$. Models like those depicted in the right panel of 
Figure \ref{fig:per} can thus be accommodated 
 even if we assume that the NSs in GW170817 had essentially no spins. 
In this way the limits 
presented in \cite{Ruiz:2017due} not only explain both events GW170817 and GW190814  
but even allow for a secondary in GW190814 that is a 
slowly or nonrotating NS. 
The end result is that we never have to resort to any exotic 
physics to explain GW190814. We also remark that the representative 
DD2 stiff EOS yields a radius of $R=13.3$ km for a mass $M=1.44~M_{\odot}$ and period 
$P=4.9$ ms, which is consistent with the results of NICER. This model resides slightly above the TOV
curve for this EOS (green star in Figure \ref{fig:per}).

The fact that rapid rotation is not necessary to explain the lighter object in GW190814 is also consistent 
with the dimensionless spin diagnostics \citep{Abbott:2020khf}. 
Assuming that $m_1$ corresponds to the BH and $m_2$ to the NS the 
effective inspiral spin parameter $\chi_{\rm eff}$ is 
\begin{equation}
\chi_{\rm eff} = \frac{\chi_1+ q\chi_2}{1+q} \leq 0.063 + 0.1\chi_2 \, ,
\label{eq:chi}
\end{equation} 
where $q=m_2/m_1=0.112$ and $\chi_1\leq 0.07$ from \citep{Abbott:2020khf}. 
Given that $\chi_{\rm eff}=-0.002^{+0.060}_{-0.061}$ \citep{Abbott:2020khf}, Equation (\ref{eq:chi})
yields $\chi_2\geq -0.05$ or $\chi_2\geq-1.26$, both of which accommodate nonrotating NSs.

Finally, we recognize that alternative models for the secondary in 
GW190814 include low-mass BHs \citep{Gupta:2019nwj},
or even an accreting NS in a circumbinary accretion disk \citep{Safarzadeh:2020ntc}.
Several viable formation scenarios      
exist for  $2.6 M_{\odot}$ BHs, such as binary NS or 
NS-white dwarf coalescence \citep{Paschalidis:2010dh,Paschalidis:2011ez}, 
whose merger remnants may collapse to form BHs in both
cases. The key point is that explaining
the secondary in GW 190814, whether as an NS or a BH,
does not require unconventional or exotic physics \citep{Vattis:2020iuz}, although such
a possibility cannot be ruled out.


\acknowledgements

It is a pleasure to thank G. Baym and N. Yunes for useful discussions.
This work was supported by National Science Foundation grant No. PHY-1662211 and No. PHY-2006066, and
the National Aeronautics and Space Administration (NASA) grant No. 80NSSC17K0070 to the
University of Illinois at Urbana-Champaign. This work made use of the
Extreme Science and Engineering Discovery Environment, which is supported by National
Science Foundation grant No. TG-MCA99S008. This research is part of the Blue Waters
sustained-petascale computing project, which is supported by the National Science Foundation
(grants No. OCI-0725070 and No. ACI-1238993) and the State of Illinois. Blue Waters
is a joint effort of the University of Illinois at Urbana-Champaign and its National Center
for Supercomputing Applications. Resources supporting this work were also provided by the
NASA High-End Computing Program through the NASA Advanced  Supercomputing Division at Ames
Research Center.

\bibliographystyle{aasjournal}       
\bibliography{references}            

\end{document}